\documentclass[fleqn,10pt]{wlscirep}
\usepackage[utf8]{inputenc}
\usepackage[T1]{fontenc}
\usepackage{lineno}
\usepackage{tabularx}%Añadido por mi
\usepackage{multirow}
\usepackage{float} %AÑADIDO POR MI
\usepackage{placeins}

\usepackage{booktabs}

\usepackage{listings}
\usepackage{xcolor}

\lstdefinestyle{directorytree}{
    basicstyle=\ttfamily\small,
    columns=fullflexible,
    keepspaces=true,
    showstringspaces=false,
    breaklines=true,
    frame=single,
    framerule=0.3pt,
    rulecolor=\color{black!30},
    backgroundcolor=\color{black!3},
    xleftmargin=0.5em,
    xrightmargin=0.5em
}

\title{A Multimodal Dataset of Student Oral Presentations with Sensors and Evaluation Data}

\author[1,*]{Alvaro Becerra}
\author[1]{Ruth Cobos}
\author[1]{Roberto Daza}

\affil[1]{Universidad Autonoma de Madrid, School of Engineering, Madrid, 28049, Spain}

\affil[*]{Corresponding author: Alvaro Becerra (alvaro.becerra@uam.es)}

\begin{abstract}
Oral presentation skills are a critical component of higher education, yet comprehensive datasets capturing real-world student performance across multiple modalities remain scarce. To address this gap, we present SOPHIAS (\textbf{S}tudent \textbf{O}ral \textbf{P}resentation monitoring for \textbf{H}olistic \textbf{I}nsights \& \textbf{A}nalytics using \textbf{S}ensors), a 12-hour multimodal dataset containing recordings of 50 oral presentations delivered by 65 undergraduate and master's students at the Universidad Autónoma de Madrid, comprising 46 individual presentations with a mean presentation duration of 9 min 48 s ($SD = 33$ s) followed by a mean Q\&A duration of 6 min 28 s ($SD = 3$ min 02 s), and 4 group presentations with a mean presentation duration of 14 min 11 s ($SD = 1$ min 40 s) followed by a mean Q\&A duration of 8 min 22 s ($SD = 1$ min 21 s). SOPHIAS integrates eight timestamped sensor streams from high-definition webcams, ambient and webcam audio, eye-tracking glasses, smartwatch physiological sensors, and clicker, keyboard and mouse interactions. In addition, the dataset includes slides and rubric-based evaluations from teachers, peers, and self-assessments, along with timestamped contextual annotations. The dataset captures presentations conducted in real classroom settings, preserving authentic student behaviors, interactions, and physiological responses. SOPHIAS enables the exploration of relationships between multimodal behavioral and physiological signals and presentation performance, supports the study of peer-assessment and provides a benchmark for developing automated feedback and Multimodal Learning Analytics tools. The dataset is available for research, including both academic and legitimate commercial research and development, under controlled access through Science Data Bank, subject to approval of a Data Usage Agreement (DUA), with code provided through GitHub.
\end{abstract}

\begin{document}

\flushbottom
\maketitle

\thispagestyle{empty}

%\noindent Please note: Abbreviations should be introduced at the first mention in the main text – no abbreviations lists or tables should be included. Structure of the main text is provided below.

\section*{Background \& Summary}

%(700 words maximum) An overview of the study design, the assay(s) performed, and the created data, including any background information needed to put this study in the context of previous work and the literature. The section should also briefly outline the broader goals that motivated the creation of this dataset and the potential reuse value. We also encourage authors to include a figure that provides a schematic overview of the study and assay(s) design. The Background \& Summary should not include subheadings. This section and the other main body sections of the manuscript should include citations to the literature as needed. 

Oral presentation skills constitute a core academic and professional competence for university students. The ability to communicate ideas clearly, confidently, and persuasively is essential for students’ future careers across disciplines. Despite their importance, many higher-education learners continue to struggle with structuring their messages, maintaining audience engagement, and managing anxiety during public speaking. Structured instruction, guided practice, and explicit rubric-based feedback are therefore important for developing strong presentation skills \cite{vzivkovic2014importance}.

Different approaches and research efforts have focused on investigating how to help students improve their oral communication skills. For example, prior studies\cite{miskam2020using, tailab2020use} indicate that video-based instructional strategies can significantly enhance students’ presentation performance. By reviewing their recorded presentations, learners can analyze aspects such as content development, fluency, pronunciation, vocabulary range, and overall organization, comparing their delivery with performance standards and engaging in targeted self-correction \cite{miskam2020using}. Video technology also promotes deeper self-reflection and helps students develop greater awareness of their strengths and weaknesses, thereby contributing to more effective and confident oral communication \cite{tailab2020use}.

Peer assessment \cite{viberg2024exploring} also plays a significant role in developing oral presentation skills. Studies show that although individual peer ratings may be less reliable than teacher assessments, aggregating several peer scores produces substantially higher reliability. Magin et al \cite{magin2001peer} reported that two to four peer ratings can match the reliability of a single teacher rating, while De Grez et al \cite{de2012effective} demonstrated that well-designed rubrics and repeated assessment opportunities yield acceptable validity and support students’ learning through reflective evaluation.

However, recent advances in Multimodal Learning Analytics (MMLA) \cite{ouhaichi2023research, ouyang2026systematic,giannakos2022multimodal} highlight the importance of integrating multiple data streams \cite{chango2022review, becerra2023m2lads, mu2020multimodal}, such as video, audio, or interaction logs, to obtain a richer and more holistic understanding of students’ performance \cite{giannakos2022multimodal, becerra2025enhancing_review, zerkouk2025predicting, mihoubi2025sentidrop, li2025and}. Physiological signals have been shown to provide meaningful indicators of attention and distraction \cite{becerra2025ai, daza2024mebal2}, motivation \cite{sharma2020eye}, learners’ activity recognition \cite{srivastava2018combining, navarro2024vaad}, recognition of sign language \cite{computers13020037}, nervousness and stress \cite{pijeira2025electrodermal}, emotion recognition \cite{liu2024eeg}, or cognitive load \cite{li2024measuring, golrang2025does}. Beyond education, multimodal behavioral cues such as eye gaze, facial expressions, attention changes, and verbal patterns are also being investigated in AI-assisted behavioral and nonclinical assessment contexts \cite{bidwe2025nonclinical}.

Recent developments in MMLA have led to the emergence of automated feedback systems designed to support learners during oral presentations. For example, Ochoa et al \cite{ochoa2024openopaf} describe how modern MMLA-based tools integrate video, audio, posture, and interaction logs to generate real-time feedback on delivery, engagement, and slide usage, illustrating a growing trend toward data-driven support for communication skills. However, as highlighted by Hummel et al \cite{hummel2025enhancing}, most existing systems are still evaluated with very small samples, often involving few presenters and tested in controlled laboratory environments rather than real classrooms, which limits their external validity and their applicability to real educational settings.

\begin{figure}
    \centering
    \includegraphics[width=\linewidth]{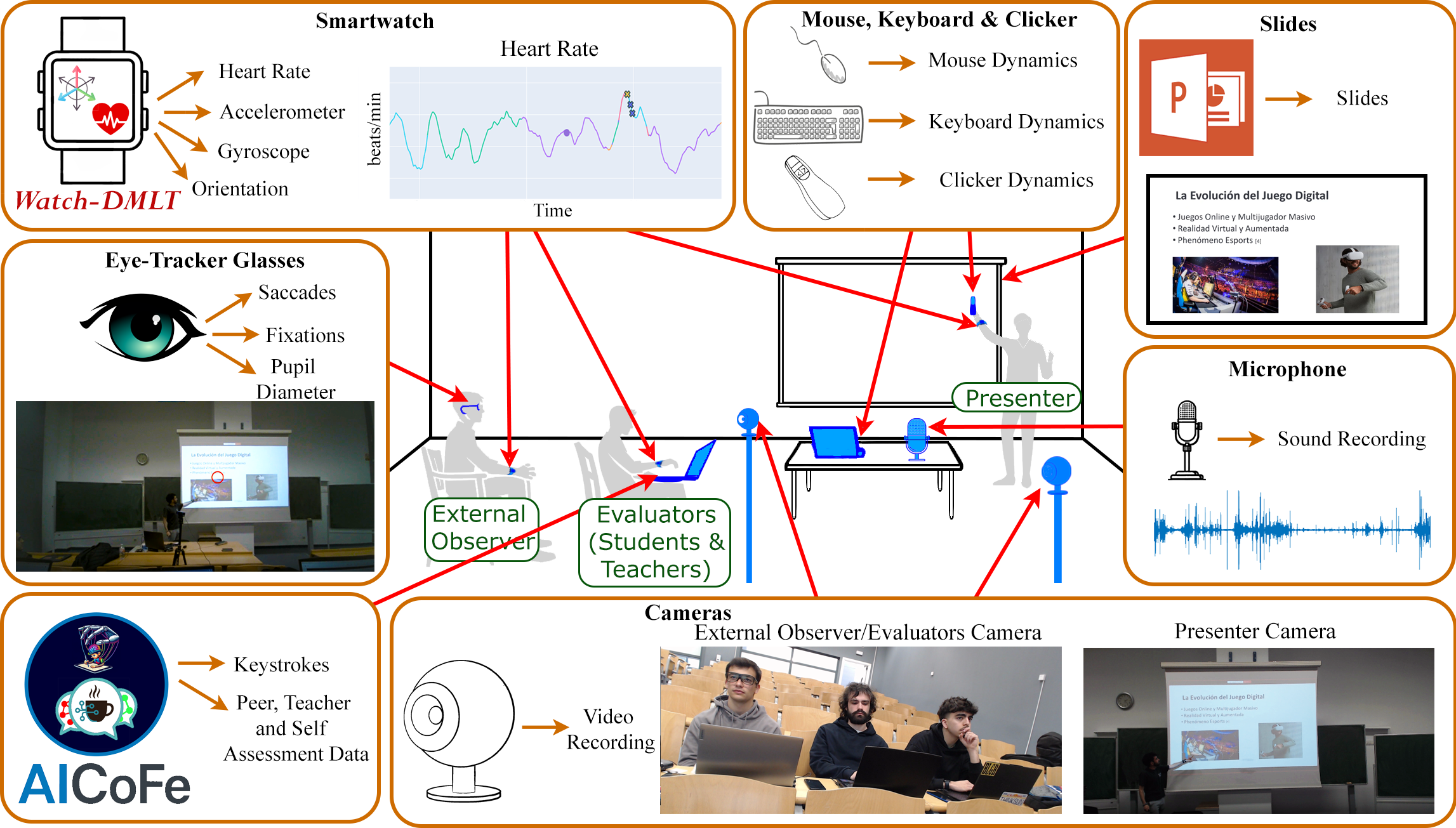}
\caption{Schematic overview of the experimental classroom setup and the synchronized sensing infrastructure deployed during student oral presentations. The figure illustrates all sensing devices and data sources used during data collection: Fitbit Sense smartwatches worn by the presenter, peer evaluators, teacher, and an external observer, capturing physiological and motion data such as heart rate and inertial signals at 1 Hz; Tobii Pro Glasses 3 eye-tracking glasses worn by the external observer, providing gaze-related measures at 50 Hz, first-person video at 25 Hz and 1920$\times$1080 resolution, and inertial measurement unit (IMU) data at 120 Hz; two Logitech C920 PRO HD external webcams recording the presenter view and the evaluators' view at 20 Hz and 1920$\times$1080 resolution; ambient audio recorded with a Poly Sync 40 microphone at 16 kHz and webcam-based audio recordings at 8 kHz; mouse, keyboard, and clicker interaction logs; presentation materials in PDF format; contextual event annotations; and rubric-based teacher, peer, and self-assessments collected through the AICoFe (\textit{AI-based Collaborative Feedback}) system. The spatial arrangement of the presenter, peer evaluators, teacher, and external observer is depicted, together with representative examples of the multimodal data streams captured from each modality. The participants provided consent for the publication of these images.}
\label{fig:setup}
\end{figure}

To address this gap, we introduce SOPHIAS (\textbf{S}tudent \textbf{O}ral \textbf{P}resentation monitoring for \textbf{H}olistic \textbf{I}nsights \& \textbf{A}nalytics using \textbf{S}ensors), a 12-hour multimodal dataset capturing real-world oral presentations from 65 undergraduate and master’s students at the Universidad Autónoma de Madrid. The dataset integrates synchronized streams from three platforms and eight sensors including high-definition webcams, ambient audio, eye-tracking glasses, smartwatch physiological signals, PDF slides, clicker/keyboard/mouse interactions, rubric-based evaluations from teachers, peers, and self-assessment, and timestamped contextual annotations (see Figure \ref{fig:setup}). 

SOPHIAS provides real recordings collected in natural classroom conditions, preserving the spontaneity, pressure, and interaction patterns inherent to real presentations, elements that are typically absent in lab-based studies. Its multimodal synchronization allows fine-grained exploration of relationships between biometric and behavioral signals. In addition, the dataset includes assessments from three sources (teachers, peers, and self-assessments) providing complementary evaluative perspectives within the same presentation sessions.

In total, the dataset comprises 385 GB of data and includes 12 hours of presenter recordings collected during oral presentation sessions. It contains 50 presentations in total, comprising 46 undergraduate individual presentations and 4 master’s group presentations, offering substantial variability in topics, delivery styles, group coordination patterns, and audience interaction.

As shown in Table~\ref{tab:related_datasets}, SOPHIAS complements existing public MMLA and educational multimodal datasets by focusing on real face-to-face oral presentations and by combining synchronized multimodal sensing, slides, interaction logs, timestamped contextual annotations, and teacher, peer, and self-assessments within the same classroom-based dataset.

\begin{table*}[t]
\centering
\scriptsize
\setlength{\tabcolsep}{3pt}
\renewcommand{\arraystretch}{1.2}
\caption{Overview of SOPHIAS and related multimodal learning analytics and educational multimodal datasets.}
\label{tab:related_datasets}
\begin{tabularx}{\textwidth}{|
>{\raggedright\arraybackslash}p{3.0cm}|
>{\raggedright\arraybackslash}p{2.1cm}|
>{\raggedright\arraybackslash}p{1.8cm}|
>{\raggedright\arraybackslash}X|
>{\raggedright\arraybackslash}X|
}
\hline
\textbf{Dataset / resource} & \textbf{Type} & \textbf{Participants} & \textbf{Modalities} & \textbf{Evaluation / annotations} \\
\hline

Math Data Corpus~\cite{oviatt2013multimodal} 
& MMLA dataset for collaborative problem solving 
& 18 students; 12 sessions 
& Speech, digital pen input, images/video 
& Problem segmentation, correctness, and written representation annotations \\
\hline

Public Speaking / Oral Presentation Quality Corpus~\cite{chen2014using} 
& Public-speaking / oral-presentation corpus 
& 17 speakers 
& Audio, video, slides, posture, hand gestures, head-pose and speech-derived features 
& Human scores related to presentation delivery and slide quality \\
\hline

mEBAL / mEBAL2~\cite{daza2020mebal,daza2024mebal2} 
& Multimodal e-learning / attention dataset 
& 38 students in mEBAL; 180 students in mEBAL2 
& RGB and NIR facial images/videos, eye-blink events, EEG/cognitive activity signals 
& Eye-blink and attention-level labels \\
\hline

IMPROVE~\cite{daza_multimodal_2025} 
& Multimodal online education dataset 
& 120 learners 
& Video, EEG, eye tracking, smartwatch data, keystroke dynamics, mobile-phone interaction, academic performance data 
& Mobile-phone use, learning-related variables, and performance measures \\
\hline

Lecture Presentations Multimodal Dataset~\cite{lee2023lecture} 
& Educational video / slide-language benchmark 
& 10 lecturers; 180+ hours of video 
& Lecture slides, spoken language, figures, transcripts 
& Figure-to-text and text-to-figure retrieval annotations \\
\hline

\textbf{SOPHIAS }
& Ecological MMLA dataset for oral presentations 
& 65 students; 50 presentations 
& HD classroom video, ambient and webcam audio, eye-tracking, smartwatch physiological/inertial signals, clicker/keyboard/mouse logs, slides, processed visual features 
& Timestamped contextual annotations; teacher, peer, and self-assessment rubrics \\
\hline

\end{tabularx}
\end{table*}

\section*{Methods}

%The Methods should include detailed text describing any steps or procedures used in producing the data, including full descriptions of the experimental design, data acquisition assays, and any computational processing (e.g. normalization, image feature extraction). See the detailed section in our submission guidelines for advice on writing a transparent and reproducible methods section. Related methods should be grouped under corresponding subheadings where possible, and methods should be described in enough detail to allow other researchers to interpret and repeat, if required, the full study. Specific data outputs should be explicitly referenced via data citation (see Data Records and Citing Data, below).

%Authors should cite previous descriptions of the methods under use, but ideally the method descriptions should be complete enough for others to understand and reproduce the methods and processing steps without referring to associated publications. There is no limit to the length of the Methods section. Subheadings should not be numbered.

\subsection*{Ethics and Consent}

The SOPHIAS dataset was collected in accordance with the Declaration of Helsinki and was approved by the Ethics Committee of the Universidad Autónoma de Madrid (Approval No. CEI-155-3850). The study was also conducted within the framework of the teaching innovation project EPS\_021.24\_INN.

Before data collection, all participants received information about the purpose of the study, the classroom data-collection procedure, the sensors and recording devices involved, the types of data to be collected, and the intended research use of the dataset. Each participant signed an informed consent form prior to participation. The consent procedure allowed students to choose their level of participation, including whether they agreed to wear specific sensors and whether they agreed to be recorded through audio and video. Participation, partial participation, or refusal to participate had no effect on students’ course grades or academic assessment.

Because SOPHIAS contains human-subject data, including faces, voices, gaze information, physiological signals, demographic variables, assessment scores, and textual comments, the dataset was not treated as fully anonymised. Instead, direct institutional identifiers were replaced by random internal identifiers generated through the AICoFe (\textit{AI-based Collaborative Feedback}) system \cite{becerra2025enhancing}. These identifiers are used to preserve the structure of the dataset and enable research interpretation while avoiding the release of students’ institutional identities.

Raw audiovisual recordings still contain identifiable information, including facial appearance, voice, body posture, gestures, and contextual classroom information. Facial blurring was not applied to the released MP4 recordings because facial expressions, gaze direction, head pose, and body posture are core research signals for multimodal learning analytics and automated feedback research. Participants provided explicit consent for the use of these recordings in academic publications and for controlled research reuse under the terms of the Data Usage Agreement.

Furthermore, participant names detected in the audio were censored by replacing the corresponding intervals with silence while preserving the original timing of the recordings, as described in Data Processing. Access to the full dataset is governed by a controlled-access model and a Data Usage Agreement, which prohibits redistribution of the original data, extracted features, encrypted copies, or derived data that could compromise participant confidentiality. Users are also required to avoid re-identification attempts and to ensure that any publications, examples, models, or released outputs do not reveal participant identities (see Data Access and Usage Notes).

Any request for consent withdrawal will be reviewed to determine the appropriate removal of the participant’s data while maintaining the coherence and consistency of the dataset.

\subsection*{Dataset: SOPHIAS} 

The dataset consists of recordings from 65 students enrolled in face-to-face education at the Universidad Autónoma de Madrid (UAM), who delivered oral presentations either individually or in groups. In total, 46 undergraduate students (40 male, 6 female), enrolled in the final year of the Telecommunication Technology and Service Engineering degree, gave individual presentations as part of the course Engineering and Society, which requires students to prepare and deliver a presentation on engineering-related topics. Individual presentations had a mean presentation duration of 9 min 48 s ($SD = 33$ s; range $= 8$ min 08 s--10 min 38 s), followed by a mean question-and-answer duration of 6 min 28 s ($SD = 3$ min 02 s; range $= 2$ min 13 s--14 min 41 s) . The average age of undergraduate students was $21.6$ years ($SD = 0.84;$ range $= 21\text{-}25$).

Additionally, 19 master’s students (14 male, 5 female) from the Master in Data Science participated in group presentations within the course Research Project in Data Science, in which students write an academic paper and subsequently present it. These students formed groups of 4 or 5 students. Group presentations had a mean presentation duration of 14 min 11 s ($SD = 1$ min 40 s; range $= 12$ min 54 s--16 min 38 s), followed by a mean question-and-answer duration of 8 min 22 s ($SD = 1$ min 21 s; range $= 6$ min 45 s--9 min 29 s). The average age of master's students was $23.6$ years ($SD =1.26;$ range $= 22\text{-}27$).

Overall, SOPHIAS contains 50 distinct presentation sessions: 46 individual undergraduate presentations and 4 master's group presentations. Thus, the number of students exceeds the number of presentations because master's students presented in groups rather than individually. Each presentation session corresponds to one presentation folder and one presentation topic.

\subsection*{Protocol} 
Prior to data collection, undergraduate and master’s students participated in two preparatory sessions.
The first session focused on oral presentation skills and included a formative instructional workshop delivered by the course teacher, who introduced key guidelines, strategies, and best practices for preparing and delivering effective presentations.

The second session introduced the teaching innovation project, explaining the implementation of peer- and self-assessment methodologies and the integration of wearable sensors and multimodal recording devices. The aim of this setup was to collect objective behavioral and physiological data to support the deployment of the MOSAIC-F framework \cite{becerra2025mosaic}, a data-driven model that combines Multimodal Learning Analytics, Observations, Sensors, Artificial Intelligence (AI), and Collaborative assessments to generate personalized feedback. During this session, students also received dedicated training on the AICoFe  system \cite{becerra2025enhancing} (see the Data, Platforms and Sensors subsection for platform details), including a detailed explanation of the evaluation rubric and its level descriptors to ensure consistency and reliability across evaluators. At the end of this session, all participants were provided with an informed consent form outlining the data collection procedures and were allowed to choose their preferred level of participation.

Following these two preparatory sessions, students delivered their oral presentations as part of the data collection process. After all presentations had been completed, students undertook a post-test to measure learning gains related to effective oral presentation skills.

Undergraduate presentations were conducted individually and took place between March and May 2025. Each student was randomly assigned an engineering-related topic (e.g., Engineering Advances 1750–1900, Robotics, or Cybersecurity and Cybercrime). For each presentation, two students were randomly selected to act as peer evaluators, assessing the presenter along with the teacher using the standardized rubric in AICoFe. Additionally, one student assumed the role of external observer and, following proper calibration, wore the Tobii Pro Glasses 3 eye-tracking device (see the Data, Platforms and Sensors subsection for sensor details). A research assistant simultaneously annotated contextual events during each presentation using a dedicated annotation interface. These annotations included nervous movements, reading behavior, and instances of eye contact with the audience. After completing their presentations, students also carried out a self-assessment using the same AICoFe rubric, reviewing the video recording of their own performance. Once the self-assessment was submitted, students gained access to personalized feedback automatically generated by GePeTo \cite{becerra2024generative}, an AI-based feedback module.

For data organization and analysis, each participant involved in an undergraduate presentation was assigned a specific role within the session. These roles included the Presenter (P), the Teacher evaluator (T), an External observer (O), and two Peer evaluators (E1 and E2). These role identifiers were used consistently throughout the dataset to distinguish the contributions and actions of each participant during the presentation and evaluation process.

For the master’s students, the protocol followed a similar structure, but with several key differences. Students selected their own topic within the field of Data Science and were required to first write a scientific article on that topic before presenting it orally. Examples of selected topics included \textit{Prediction and Early Warning of Fetal Hypoxia} and \textit{Optimization of Railway Transportation Networks}. All master’s presentations took place on the same day in May 2025.

During the master’s session, both teachers and students used the AICoFe system to conduct their evaluations. Each teacher was assigned one student to evaluate individually, focusing on specific dimensions such as vocal delivery or body expression. In addition to these individual assessments, teachers also evaluated group-level aspects of the presentations, including slide design, the conclusions, and the coordination among team members. Similarly, each master’s student evaluated one peer using the same rubric, assessing both individual dimensions and group-level aspects.

Due to the group-based nature of the master’s presentations, the role naming scheme was extended to accommodate multiple presenters and evaluators. Presenter roles were indexed as P1–P5, peer evaluators as E1–E5, and teachers as T1–T5, while the external observer was consistently identified as O.

\begin{table}[h!]
\centering
\begin{tabular}{|l|c|c|l|l|}
\hline
\textbf{Degree} & \textbf{ID Item} & \textbf{ID Section} & \textbf{Section Name} & \textbf{Item Description} \\
\hline\hline

% =======================
% GRADO (36–50)
% =======================
\multirow{15}{*}{Undergraduate}
 & 36 & \multirow{4}{*}{1} & \multirow{4}{*}{Content} & Opening of the presentation \\ \cline{2-2}\cline{5-5}
 & 37 &  &  & Content development \\ \cline{2-2}\cline{5-5}
 & 38 &  &  & Closing of the presentation \\ \cline{2-2}\cline{5-5}
 & 39 &  &  & Bibliography and references \\ \cline{2-5}

 & 40 & \multirow{3}{*}{2} & \multirow{3}{*}{Design} & Text readability \\ \cline{2-2}\cline{5-5}
 & 41 &  &  & Relevance of visual elements \\ \cline{2-2}\cline{5-5}
 & 42 &  &  & Visual balance \\ \cline{2-5}

 & 43 & \multirow{8}{*}{3} & \multirow{8}{*}{Presenter Performance} & Voice and tone \\ \cline{2-2}\cline{5-5}
 & 44 &  &  & Body language \\ \cline{2-2}\cline{5-5}
 & 45 &  &  & Eye contact with the audience \\ \cline{2-2}\cline{5-5}
 & 46 &  &  & Use of complementary resources \\ \cline{2-2}\cline{5-5}
 & 47 &  &  & Ability to capture attention \\ \cline{2-2}\cline{5-5}
 & 48 &  &  & Autonomy during the presentation \\ \cline{2-2}\cline{5-5}
 & 49 &  &  & Handling of questions \\ \cline{2-2}\cline{5-5}
 & 50 &  &  & Mastery of the topic \\
\hline\hline

% =======================
% MÁSTER (51–64)
% =======================
\multirow{14}{*}{Master}
 & 51 & \multirow{5}{*}{6} & \multirow{5}{*}{Content (Group-level)} & Opening of the presentation \\ \cline{2-2}\cline{5-5}
 & 52 &  &  & Content development \\ \cline{2-2}\cline{5-5}
 & 53 &  &  & Closing of the presentation \\ \cline{2-2}\cline{5-5}
 & 54 &  &  & Bibliography and references \\ \cline{2-2}\cline{5-5}
 & 55 &  &  & Group performance \\ \cline{2-5}

 & 56 & \multirow{3}{*}{7} & \multirow{3}{*}{Design (Group-level)} & Text readability \\ \cline{2-2}\cline{5-5}
 & 57 &  &  & Relevance of visual elements \\ \cline{2-2}\cline{5-5}
 & 58 &  &  & Visual balance \\ \cline{2-5}

 & 59 & \multirow{6}{*}{8} & \multirow{6}{*}{Presenter Performance (Individual-level)} & Voice and tone \\ \cline{2-2}\cline{5-5}
 & 60 &  &  & Body language \\ \cline{2-2}\cline{5-5}
 & 61 &  &  & Eye contact with the audience \\ \cline{2-2}\cline{5-5}
 & 62 &  &  & Autonomy during the presentation \\ \cline{2-2}\cline{5-5}
 & 63 &  &  & Handling of questions \\ \cline{2-2}\cline{5-5}
 & 64 &  &  & Mastery of the topic \\
\hline

\end{tabular}
\caption{Evaluation rubric items organized by degree level, section, and description.  The Undergraduate rubric is applied exclusively to undergraduate presentations, and the Master’s rubric is applied exclusively to Master’s presentations
\emph{ID Item} denotes the unique internal identifier assigned to each rubric item within the AICoFe assessment system; \emph{ID Section} identifies the rubric section to which each item belongs and is used internally to group criteria within the evaluation platform. 
Undergraduate presentations use individual-level criteria, whereas Master’s presentations distinguish between group-level and individual-level assessment dimensions}
\label{tab:items}
\end{table}

\subsection*{Classroom Environment and Recording Setup}

The recordings were collected in a regular classroom at the Universidad Autónoma de Madrid under real teaching conditions. The room had an approximate width of 10.40 m and length of 12.90 m. The external webcam recording the presenter was positioned approximately at the centre of the classroom width and at a distance of approximately 5 m from the projection screen. The presenter stood at the front of the classroom, approximately 4.00–4.10 m from the webcam. Students were seated facing the presenter and the projection screen, following the standard classroom seating arrangement. The acoustic environment corresponded to a typical indoor classroom setting.

\subsection*{Assessment Rubric}

All assessments in both undergraduate and master's courses were conducted using a standardized analytic rubric implemented within the AICoFe system. The rubric was developed collaboratively by five teachers from the UAM, all of whom have extensive experience teaching courses on oral communication and presentation skills in engineering education. It was designed to evaluate the quality of oral presentations through multiple complementary dimensions, and within the AICoFe system the full rubric and its level descriptors were accessible at any time to all evaluators and students (see Table~\ref{tab:items}). The teaching team analyzed key competencies and expected learning outcomes while consulting relevant literature on rubric design for oral presentations \cite{van2017assessing, nadolski2021rubric, galvan2017assessing, peeters2010standardized}.

For the undergraduate students, the rubric comprised three main dimensions: \textit{Content}, \textit{Design}, and \textit{Presenter Performance}. The \textit{Content} and \textit{Design} dimensions assessed the structure, clarity, visual quality, and coherence of the presentation, while the \textit{Presenter Performance} dimension captured speaker-related aspects such as vocal delivery, body language, eye contact, engagement strategies, autonomy during the presentation, management of questions, and mastery of the topic. Each item was rated on a 5-point Likert scale (1 = Very Poor, 5 = Excellent).

The master’s rubric followed a similar structure but was adapted to the characteristics of group research presentations and the requirements of an academic paper defense. Accordingly, the rubric differentiated between \textit{group-level} and \textit{individual-level} criteria. Group-level criteria evaluated the collective quality of the presentation, including the clarity and motivation of the opening, the technical development and novelty of the contribution, the effectiveness of the closing, the accuracy of citations, the coordination among presenters, and the design of the visual material. Individual-level criteria focused on each student's delivery, including voice and tone, body language, eye contact, autonomy, handling of questions, and mastery of the assigned research content. As in the undergraduate rubric, all items were assessed on a 5-point Likert scale with explicit level descriptors.

The use of a unified scoring structure across peer-, teacher-, and self-assessments ensures that the resulting dataset includes complementary evaluative perspectives while maintaining comparability across sources.

\subsection*{Data, Platforms and Sensors}\label{sec:data_platforms_sensors}
Three complementary platforms were used to monitor learners during the oral presentations: edBB (\textit{Biometrics and Behavior for Assessing Remote Education}) \cite{daza2023edbb}, Microsoft Teams, and the AICoFe system. In addition, smartwatch data were collected using the Watch-DMLT (\textit{Data Monitoring and Logging Tool for Smartwatches enabling real-time synchronization and activity data storage}) \cite{becerra2025real} acquisition tool. Each platform contributed different modalities, enabling a synchronized multimodal dataset combining physiological, behavioral, visual, audio, and evaluative information.

To facilitate synchronization across modalities, timestamped files include time information in local time and UNIX time. Local timestamps are reported in the Europe/Madrid time zone, corresponding to the time zone in which the classroom recordings were collected. UNIX timestamps are expressed in milliseconds since the UNIX epoch (1970-01-01 00:00:00 UTC). These UNIX timestamps are timezone-independent and were used as the common temporal reference for aligning multimodal streams across platforms and sensors.

The edBB platform served as the primary infrastructure for capturing audiovisual signals. It recorded audio directly from the external webcams, video from the same webcams, and screen recordings of the Teams meeting during the  presentation. All streams were timestamped and stored using edBB’s unified timing mechanism, ensuring precise synchronization across sensors.

Microsoft Teams was used to capture the high-quality audio from an unobtrusive microphone.

Smartwatch data were collected using Watch-DMLT, a real-time acquisition tool developed specifically for Fitbit Sense devices, enabling high-frequency access to physiological and motion signals that are normally restricted in commercial wearables. Watch-DMLT supported parallel acquisition from multiple watches worn by presenters, teachers, evaluators, and observers.

The AICoFe system was used to manage all evaluations during the sessions. It recorded rubric scores, textual comments, and interaction logs (keystrokes), generating time-aligned metadata that reflect the evaluative decision processes of teachers and peers.

A multimodal acquisition framework was designed, incorporating multiple information sources and sensors (see Figure~\ref{fig:setup} and Table \ref{table:sophias}):
\begin{table}[t]
\centering
\begin{tabularx}{\textwidth}{|>{\hsize=0.9\hsize}X|>{\hsize=0.9\hsize}X|>{\hsize=0.6\hsize}X|>{\hsize=1.6\hsize}X|}
\hline
\textbf{Information Type} & \textbf{Sensors or Platforms} & \textbf{Sampling Rate} & \textbf{Data / Features} \\
\hline \hline

Video &
2 Logitech C920 PRO HD \mbox{webcams} &
20 Hz &
MP4 recordings (H.264) \\
\hline

Ambient Audio &
Poly Sync 40 microphone &
16 kHz &
WAV audio \\
\hline

Webcam Audio &
Integrated webcam microphone &
8 kHz &
WAV audio \\
\hline

Eye-tracking &
Tobii Pro Glasses 3 &
50 Hz (gaze), 25 Hz (video), 120 Hz (IMU) &
First-person video; 3D gaze vectors; pupil diameter; blink events; 6-axis IMU signals \\
\hline

Smartwatch &
Fitbit Sense &
1 Hz &
Heart rate (PPG); accelerometer; gyroscope; orientation \\
\hline

Presenter Interaction Events &
Clicker, keyboard, mouse &
Event-based &
Timestamped actions: mouse moves/clicks, keystrokes, slide navigation \\
\hline

Performance Assessment &
AICoFe system &
N/A &
Rubric-based teacher, peer, and self-evaluation; textual comments; keystroke logs \\
\hline

Contextual Annotations &
Dedicated annotation interface &
N/A &
Timestamped labels: nervous movements, reading behavior, eye contact, Q\&A events \\
\hline

Presentation Material &
PDF slide files &
N/A &
Slides \\
\hline

Face position, Head Pose and Facial Landmarks &
RetinaFace + WHENet pipeline &
20 Hz &
Face bounding boxes; five landmarks; head pose angles (yaw, pitch, roll) \\
\hline

\end{tabularx}
\caption{\label{table:sophias} Summary of sensors and data collection from the SOPHIAS dataset}
\end{table}

\begin{itemize}
    \item \textbf{Cameras:} Two external webcams were used to record the classroom:
    \begin{itemize}
        \item Presenter Camera (external): A Logitech C920 PRO HD webcam was positioned in front of the presenter to capture a frontal RGB view of the entire presentation, operating at 20 Hz with a resolution of 1920 × 1080.
        \item Evaluators/Observer Camera (external): A second Logitech C920 PRO HD webcam was oriented towards the audience area where the evaluators and the external observer were seated, operating at 20 Hz with a resolution of 1920 × 1080. 
    \end{itemize}
All video streams were encoded in standard MP4 format  (H.264 codec), and edBB recorded the exact capture timestamp for every frame, storing both UNIX and local time to enable synchronization across all sensors. The audio captured by the Logitech C920 PRO HD webcams was recorded at a sampling rate of 8 kHz.

\item \textbf{Ambient microphone:} A Poly Sync 40 omnidirectional speakerphone was positioned on the table, close to the presenter, and was connected to Microsoft Teams as the main audio input device. This non-intrusive microphone records the full audio of the session at a 16 kHz sampling rate.
\item \textbf{Fitbit Sense Smartwatches:} The presenters, the peer evaluators, and the external observer wore a Fitbit Sense smartwatch during each presentation. This device includes a photoplethysmography (PPG) sensor to estimate heart rate, a 3-axis accelerometer and 3-axis gyroscope to capture linear and rotational motion, an orientation sensor that reports device position. Using Watch-DMLT, all smartwatch signals were recorded at 1 Hz and exported as structured, timestamped files, enabling reliable alignment and synchronization with other sensors. Heart rate is an important physiological marker that reflects the body's response to various emotional and cognitive states \cite{malmberg2021revealing}. Elevated heart rates are often observed during periods of stress, anxiety, or physical exertion, while lower heart rates are typically associated with relaxation and calmness \cite{kim2018stress, choi2009using}.

\item \textbf{Eye-tracking glasses:}
One student acted as an external observer and wore Tobii Pro Glasses 3, a lightweight, head-mounted eye-tracking system designed for real-world research scenarios. The glasses use a binocular corneal-reflection, dark-pupil technique with stereo geometry to estimate gaze and eye orientation. The device captures binocular eye-tracking data at 50 Hz, including 3D gaze direction vectors, pupil diameter measurements, and eye-blink detection. In addition to the eye-tracking data, the glasses record a first-person scene video from a front-mounted camera at a resolution of 1920×1080 pixels and 25 fps, accompanied by 16-bit mono audio captured by the integrated microphone.

\item \textbf{Slides:}  The original presentation materials were collected in PDF format, corresponding to the slides used by each presenter. From these files, structural and design features can be extracted.

\item \textbf{Clicker, mouse, and keyboard events:}
Presenters advanced their slides using a wireless clicker, the mouse, or keyboard shortcuts. Since the clicker simply sent its commands in the form of keyboard-like keystroke events, both keyboard and clicker interactions were captured through the same logging mechanism. All interaction events were recorded with precise timestamps, including mouse movements, mouse clicks, and any keypress generated either by the physical keyboard or by the clicker (e.g., arrow keys, spacebar, PageUp/PageDown).

\item \textbf{Assessment Data:}
The AICoFe system was used by teachers, peer evaluators, and presenters (for self-assessment) to complete a standardized rubric. For each presentation, AICoFe stores the score assigned to each rubric item and the free-text comments provided for that item by each assessment source (teacher, peer, and self). In addition, the system records keystroke events generated during the assessment process, preserving the exact timestamps of each typed character. These keystroke logs provide fine-grained temporal information about evaluators’ commenting activity.

\item \textbf{Contextual Event Annotations:}
During each presentation, a research assistant used a dedicated annotation interface to mark contextual events in real time. The annotated events include nervous movements, episodes of reading directly from the slides or from written notes, explicit moments of eye contact with the audience, filler words, or the start and end timestamps of each individual question from the audience and each corresponding response from the presenter.

These annotations are stored with exact timestamps and later aligned with the sensor streams, providing high-quality labels for supervised learning and for validating automatic detections derived from video, audio, heart rate, or gaze data.

\end{itemize}
\begin{table}[th]
\centering
\renewcommand{\arraystretch}{1} % Ajusta el espacio entre las filas
\begin{tabularx}{\textwidth}{|X|X|X|p{1cm}|X|}
\hline
\textbf{Folder Name} & \textbf{Files} & \textbf{Description} & \textbf{Format} & \textbf{File Contents} \\
\hline
\textbf{PresenterVideo} & webcam\newline WebcamCapture & Frontal presenter webcam recording & .csv\newline .mp4 & Frame number and timestamps recorded in both UNIX and local time \\
\hline
\textbf{EvaluatorsVideo} & webcam\newline WebcamCapture & Evaluator and observer webcam recording & .csv\newline .mp4 & Frame number and timestamps recorded in both UNIX and local time \\
\hline
\textbf{SoundCapture} & Record & Session audio recorded via webcam-integrated microphone & .wav & -- \\
\hline
\textbf{TeamsAudio} & Record & Session audio recording by ambient microphone using Microsoft Teams & .wav & -- \\
\hline
\end{tabularx}
\caption{Description of files contained in the edBB folder}
\label{tab:edbb}
\end{table}

\begin{table}[th]
\centering
\setlength{\tabcolsep}{6pt} % Ajusta el espacio entre las columnas
\renewcommand{\arraystretch}{1} % Ajusta el espacio entre las filas
\begin{tabularx}{\textwidth}{|X|X|X|p{1cm}|X|}
\hline
\textbf{Folder Name} & \textbf{Files} & \textbf{Description} & \textbf{Format} & \textbf{File Contents} \\
\hline
\textbf{Role} (e.g., P, T, O, T1, T2, etc.) & 
HeartRate \newline Gyroscope \newline Accelerometer \newline Orientation & 
 Fitbit Sense data & 
.csv \newline .csv \newline .csv \newline .csv & 
Acceleration on X, Y, Z axes (m/s\(^2\)), angular velocity on X, Y, Z axes (\(^\circ/s\)), heart rate (bpm), scalar part $w$ and vector parts $x,y,z$ representing device orientation, local time and UNIX timestamp \\
\hline
\textbf{Role\_filter} & 
HeartRate \newline Gyroscope \newline Accelerometer & 

Fitbit Sense data filtered using a Butterworth low-pass filter or a moving average filter& 
.csv \newline .csv \newline .csv & 
Acceleration on X, Y, Z axes (m/s\(^2\)), angular velocity on X, Y, Z axes (\(^\circ/s\)), heart rate (bpm), local time and UNIX timestamp \\
\hline
\end{tabularx}
\caption{Example of smartwatch data contained in the Watch-DMLT folder. Each participant role has its own subfolder, and all roles in both individual and group presentations (e.g., P, T, O, E1, E2) follow the same file structure and include the same set of physiological and motion signals. The folder contains both raw signals and preprocessed versions of the same signals}
\label{tab:Watch-dmlt}
\end{table}

\begin{table}[th]
\centering
\setlength{\tabcolsep}{6pt} % Ajusta el espacio entre las columnas
\renewcommand{\arraystretch}{1} % Ajusta el espacio entre las filas
\begin{tabularx}{\textwidth}{|X|X|X|p{1cm}|p{6cm}|}
\hline
\textbf{Folder Name} & \textbf{Files} & \textbf{Description} & \textbf{Format} & \textbf{File Contents} \\
\hline
\textbf{Assessments} & AICoFe presenter ID & Rubric-based assessment scores and textual comments from peers, teacher, and self-assessment & .csv & Item IDs and sections of each evaluated criterion; evaluators, teacher, and presenter AICoFe IDs; peer/teacher/self scores; and qualitative comments from each item and assessment source \\
\hline
\textbf{Slides} & slides & Slides of the presenter & .pdf & -- \\
\hline
\textbf{Logs} & 
events \newline presentation\_data \newline keys\_role (e.g., keys\_E1, keys\_T1, etc.) \newline mouse\_navigation \newline keyboard\_navigation & 
Manual annotations, AICoFe system logs, and presenter interaction data & 
.csv \newline .csv \newline .csv \newline .csv \newline .csv & 
Activity/event label, start and end timestamps, and AICoFe ID identifying the presenter (in group settings) or the participant asking a question \newline
Evaluator, teacher, observer, and presenter AICoFe IDs; theme name; smartwatch-hand assignment; and start timestamps in both local time and UNIX format \newline
AICoFe ID, key pressed, and UNIX timestamp \newline
Event type, mouse coordinates (x, y), and timestamps in both local time and UNIX format \newline
Key pressed and timestamps in both local time and UNIX format \\
\hline
\end{tabularx}
\caption{Description of files contained in the AICoFe folder}
\label{tab:AICoFe}
\end{table}

% ------------------------------
\begin{table}[th]
\centering
\setlength{\tabcolsep}{6pt}
\renewcommand{\arraystretch}{1}
\begin{tabularx}{\textwidth}{|X|X|X|p{1cm}|X|}
\hline
\textbf{Folder Name} & \textbf{Files} & \textbf{Description} & \textbf{Format} & \textbf{File Contents} \\
\hline
\textbf{PresenterWebcam} & box \newline head\_pose \newline landmarks & Processed video data: Bounding boxes, facial landmarks, and head pose of the presenter webcam & .csv \newline .csv \newline .csv & Bounding box is formatted as [x1, y1, x2, y2]; 5 landmarks per detected face; Euler angles (yaw, pitch, roll) in degrees; frames \\
\hline
\textbf{Eyetracker} & box \newline head\_pose \newline landmarks & Processed eyetracker video data: Bounding boxes, facial landmarks, and head pose of the eyetracker video & .csv \newline .csv \newline .csv & Bounding box is formatted as [x1, y1, x2, y2]; 5 landmarks per detected face; Euler angles (yaw, pitch, roll) in degrees; frames \\
\hline
\end{tabularx}
\caption{Description of files contained in the face\_processing folder}
\label{tab:face_processing}
\end{table}

% ------------------------------
\begin{table}[th]
\centering
\setlength{\tabcolsep}{6pt}
\renewcommand{\arraystretch}{1}
\begin{tabularx}{\textwidth}{|X|X|X|p{1cm}|p{8cm}|}
\hline
\textbf{Folder Name} & \textbf{Files} & \textbf{Description} & \textbf{Format} & \textbf{File Contents} \\
\hline
\textbf{Video} & scenevideo & Tobii Pro Glasses 3 video recording & .mp4 & -- \\
\hline
\textbf{Data} & gazedata \newline imudata \newline recording & Data captured by the Tobii Pro Glasses 3 & .json \newline .json \newline .json & Gaze stream, sampled at 50 Hz, including 2D and 3D gaze vectors, left/right eye origins and directions, pupil diameter, and timestamps \newline IMU (Inertial Measurement Unit) stream sampled at 120 Hz, containing 3-axis accelerometer data (m/s²) and 3-axis gyroscope data (deg/s), each entry with a precise timestamp. \newline Recording metadata containing global session information (creation time, duration, timezone, and software version) \\
\hline
\end{tabularx}
\caption{Description of files contained in the eyetracker folder}
\label{tab:eyetracker}
\end{table}

\subsection*{Data Processing} 

The frontal presenter videos and the first-person recordings from the Tobii Pro Glasses 3 were processed to obtain face detections, facial landmark tracks, and head pose estimations for each frame:
\begin{itemize}
    \item Facial detection and facial landmarks:  The facial detection module employed RetinaFace model \cite{deng2020retinaface} (\url{https://github.com/serengil/retinaface}). This robust single-stage face detector was trained using the Wider Face dataset~\cite{yang2016wider}. In SOPHIAS dataset, RetinaFace was used to automatically detect the presenter’s face in each frame and to extract five key facial landmarks (both eye centers, the nose tip, and the two mouth corners). Once the facial position in the image is obtained, it is used as input for the subsequent modules. The detector was initialized with a detection input size of 640 × 640 pixels and executed on GPU. No explicit confidence threshold was manually defined; therefore, the default threshold of the RetinaFace detector was used. For each frame, all detected faces were first clipped to the image boundaries to avoid invalid coordinates.
    \item Head Pose Estimation: A real-time head pose estimation module based on the WHENet \cite{zhou2020whenet} architecture was employed to extract the 3D orientation of the presenter’s head from 2D facial images. The module operates directly on the face crops detected by the RetinaFace model, ensuring that head pose is estimated only on accurately localized facial regions. In our implementation, EfficientNet \cite{tan2019efficientnet} served as the backbone for feature extraction, leveraging its efficient scaling strategy to capture detailed spatial features from each detected face. WHENet predicts the three Euler angles (pitch, yaw, and roll) providing a continuous characterization of head movements throughout the presentation. The architecture is based on a fully convolutional design optimized for robust angle classification and regression. The network comprises an initial block of convolutional and pooling layers, followed by fully connected layers that interpret the extracted features. Each head orientation angle is first classified into discrete bins, after which a regression stage refines the prediction within the selected bin to obtain a precise continuous value. The model was pretrained on the 300W-LP \cite{zhu2016face} and CMU Panoptic \cite{joo2015panoptic} datasets. This module was also used in the IMPROVE dataset \cite{daza_multimodal_2025}.
\end{itemize}

The audios recorded by the external webcam and the ambient microphone via Microsoft Teams were synchronized using an FFT-based cross-correlation module that computes the sample-level temporal offset between both waveforms. After loading each track, the Teams audio was resampled to match the webcam signal, normalized to reduce gain differences, and cross-correlated through convolution with its time-reversed counterpart. The lag corresponding to the global maximum of the cross-correlation function was then used to compensate for the delay by padding or trimming the Teams audio. To preserve participant privacy, both audios also underwent an automatic name-censorship process. First, the signal was transcribed using faster-whisper \cite{fasterwhisper} with word-level timestamps. A list of participant names was normalized and matched against the transcript using approximate string similarity (SequenceMatcher), allowing the detection of exact and near-matches. The detected time intervals were then merged and replaced in the waveform with silence, while preserving the original timing of the audio.

The smartwatch signals underwent a dedicated preprocessing stage to reduce noise and enhance the interpretability of physiological and motion-related measurements. Heart rate readings were smoothed using a 15-second moving average filter, an appropriate window length given that wrist-based PPG sensors capture gradual fluctuations rather than abrupt beat-to-beat changes at typical sampling rates. Inertial measurements were also filtered to suppress high-frequency artifacts. Accelerometer data were processed with a fourth-order Butterworth low-pass filter set at 0.2 Hz. This filtering attenuates rapid fluctuations and measurement noise while retaining slower wrist-motion patterns relevant to the recorded presentation activities. The gyroscope data were filtered using an equivalent Butterworth topology and cutoff frequency. These preprocessing steps were implemented in Python using pandas and NumPy for data handling and timestamp processing, and SciPy signal-processing functions, including scipy.signal.butter and scipy.signal.filtfilt, for Butterworth filter design and zero-phase forward--backward filtering. Both the unprocessed and filtered smartwatch signals are provided in the dataset.

\section*{Data Records}

The SOPHIAS dataset is hosted under controlled access in the Science Data Bank (DOI: \url{https://doi.org/10.57760/sciencedb.33655}) \cite{sophias_sdb}, while the associated code, machine-readable schemas, dataset catalog, and documentation are available through the public GitHub repository \cite{sophias_github} (\url{https://github.com/dataGHIA/SOPHIAS}). The dataset comprises 385 GB of multimodal information, including high-definition classroom videos, ambient audio, smartwatch physiological signals, eye-tracking data, interaction logs, PDF slides, and rubric-based teacher, peer, and self-assessments. These data were collected using the edBB platform, Microsoft Teams, the Watch-DMLT acquisition tool, and the AICoFe system, and include both raw and processed signals. 

To facilitate reproducible reuse and automated validation of the structured files,
we also provide a set of machine-readable JSON Schema definitions in the public
GitHub repository, under the \texttt{schemas/} directory. These schemas
document the expected fields, data types, and file-level structure for the main
tabular and JSON records. In addition, the
\texttt{dataset\_catalog.json} file maps each expected file pattern in the
SOPHIAS folder hierarchy to its corresponding schema or media-file description,
supporting consistent parsing, validation, and reproducible downstream analysis.

The dataset is organized into 50 presentation folders, each corresponding to one recorded presentation (46 undergraduate individual presentations and 4 master’s group presentations, labeled as 20250528{1--4}). Within each presentation folder, the data are structured into five main subdirectories, each reflecting a specific platform or processing stage:
\begin{itemize}
    \item edBB: Contains all audiovisual recordings captured during the presentation sessions, including the two external webcams (presenter view and evaluator view), their corresponding audio tracks, and the high-quality ambient audio recorded through Microsoft Teams (Table \ref{tab:edbb}).
        \item Watch-DMLT: Contains the physiological and motion data collected from the Fitbit Sense smartwatches via the Watch-DMLT tool, including raw and filtered heart rate, accelerometer, gyroscope, and orientation signals. Smartwatch recordings are stored following a role-based naming convention (see Methods for a detailed description of participant roles and acquisition protocol). (Table~\ref{tab:Watch-dmlt}).
    \item AICoFe: Stores the rubric-based assessment data, including teacher, peer, and self-evaluation scores, free-text comments, and the keystroke logs produced during the assessment process. This folder also contains the PDF files of the presentation slides, the clicker and keyboard interactions used to advance the slides, and the manual contextual annotations recorded by the research assistant during each presentation (Table \ref{tab:AICoFe}).
        \item face\_processing: Includes all processed visual features derived from the edBB videos, such as face detections, five facial landmarks, and head-pose estimations (Table \ref{tab:face_processing}).
    \item eyetracker: Stores the Tobii Pro Glasses 3 data, including the first-person scene video, binocular gaze samples, pupil diameter, eye-blink events, and all associated metadata (Table \ref{tab:eyetracker}).

\end{itemize}

Additionally, an extra file, \texttt{ids\_data.csv}, is included. This file specifies, for each presenter AICoFe student ID, the corresponding presentation folder in which that student acted as presenter, together with demographic information such as gender and age. AICoFe IDs are internal participant identifiers generated by the AICoFe system, whereas presentation IDs are folder-level identifiers that uniquely identify each recorded presentation session in the dataset.

Role labels indicate the function of each participant within a specific presentation session. For undergraduate individual presentations, the role labels are \texttt{P} for presenter, \texttt{T} for teacher evaluator, \texttt{O} for external observer, and \texttt{E1} and \texttt{E2} for peer evaluators. For master's group presentations, indexed role labels are used, such as \texttt{P1}--\texttt{P5} for presenters, \texttt{T1}--\texttt{T5} for teacher evaluators, \texttt{E1}--\texttt{E5} for peer evaluators, and \texttt{O} for the external observer. The mapping between AICoFe IDs and session-specific roles is provided in each presentation folder in \texttt{AICoFe/Logs/presentation\_data.csv}.

\section*{Technical Validation}

\subsection*{Dataset Completeness} Since students were allowed to choose which sensors and recordings they consented to, a validation was conducted to determine the exact amount of data available for each modality. As shown in Table \ref{tab:data_issues}, this process revealed that the final dataset contains 42 presenter view videos (4 group presenter view videos and 38 individual presenter view videos) and complete eye-tracking data are available for 39 presentations. Importantly, 37 presentations contain both presenter-view video and complete eye-tracking data. Additionally, keyboard and mouse interaction logs are available for most cases, with only a few missing due to technical issues.

The smartwatch exhibited the lowest consent rate among students, leading to fewer available recordings for this modality. Nevertheless, seven presentations contain complete smartwatch recordings for all roles involved in the session. Tables \ref{tab:available_data_group} and \ref{tab:available_data_individual} summarize the available smartwatch recordings. Considering all data sources jointly, we regard a presentation session as fully complete when all expected files are available. According to this criterion, 5 presentation sessions are fully complete.

Tables \ref{tab:data_issues}, \ref{tab:available_data_group} and \ref{tab:available_data_individual} provide complementary information on data availability and missingness and should be interpreted together. Table \ref{tab:data_issues} summarizes presentation-level missingness and technical issues affecting the camera recordings, keyboard and mouse interaction logs, and eye-tracking data. Smartwatch availability is reported separately because it depends both on participant consent and on the role performed during each session. Table \ref{tab:available_data_group} therefore summarizes role-level smartwatch availability and data-quality issues for group presentations, whereas Table \ref{tab:available_data_individual} provides the corresponding information for individual presentations. In Tables \ref{tab:available_data_group} and  \ref{tab:available_data_individual}, the absence of a presentation identifier for a student role indicates that the participant did not consent to smartwatch recording; for teacher and observer roles, it indicates that a smartwatch was not worn during that session.
\begin{table}[ht]
\centering
\begin{tabularx}{\textwidth}{|l|X|X|c|}
\hline
\textbf{Sensor / Platform} & \textbf{Data Issues} & \textbf{ID} & \textbf{Count} \\
\hline
\multirow{2}{*}{\textbf{Camera}} 
    & EvaluatorsVideo (only one evaluator consented to be recorded, so only one evaluator and the observer were recorded) 
    & 202503173, 202503181, 202503183, 202503242, 202503253, 202503311, 202503312, 202503313, 202504071, 202504073 
    & 10 \\
\cline{2-4}
    & PresenterVideo (presenter did not consent to be recorded) 
    & 202504012, 202505052, 202505053, 202505121, 202505132, 202505133, 202504082, 202504083 
    & 8 \\
\hline
\textbf{Logs} 
    & Missing keyboard\_navigation.csv and mouse\_navigation.csv 
    & 202503111, 202503242, 202504012 
    & 3 \\
\hline
\multirow{3}{*}{\textbf{Eyetracker}} 
    & Presenter did not consent to be recorded with eye-tracking glasses 
    & 202505132, 202505133, 202505121, 202505053, 202505052, 202504012, 202503172 
    & 7 \\
\cline{2-4}
    & Missing Q\&A section in recording 
    & 202503111, 202503112, 202503113 
    & 3 \\

\cline{2-4}
    & Missing first minutes in recording 
    & 202505281
    & 1 \\
\hline

\end{tabularx}
\caption{Summary of data issues detected across different sensors and platforms, including consent-related restrictions and missing recordings}
\label{tab:data_issues}
\end{table}

\begin{table}[ht]
    \centering

\begin{tabularx}{\textwidth}{|l|l|X|r|}
\hline
\textbf{Role} & \textbf{Data Issues} & \textbf{ID} & \textbf{Count} \\
\hline
\multirow{1}{*}{E2} & No issue & 202505281, 202505283, 202505284 & 3 \\
\hline
\multirow{4}{*}{E3} & No issue & 202505282 (orientation), 202505283 (accelerometer, heartRate, orientation) & 2 \\
\cline{2-4}
 & Missing accelerometer file & 202505282 & 1 \\
\cline{2-4}
 & Missing gyroscope file & 202505282, 202505283 & 2 \\
\cline{2-4}
 & Missing heartRate file & 202505282 & 1 \\
\hline
\multirow{1}{*}{P2} & No issue & 202505281, 202505282, 202505283, 202505284 & 4 \\
\hline
\multirow{3}{*}{P3} & No issue & 202505281, 202505282 (accelerometer, gyroscope, heartRate), 202505283, 202505284 (accelerometer, heartRate, orientation) & 4 \\
\cline{2-4}
 & Missing gyroscope file & 202505284 & 1 \\
\cline{2-4}
 & Missing orientation file & 202505282 & 1 \\
\hline
\multirow{1}{*}{P4} & No issue & 202505281, 202505282, 202505283, 202505284 & 4 \\
\hline
\multirow{2}{*}{P5} & No issue & 202505282, 202505283 (accelerometer, heartRate, orientation), 202505284 & 3 \\
\cline{2-4}
 & Missing gyroscope file & 202505283 & 1 \\
\hline
\multirow{1}{*}{T1} & No issue & 202505282, 202505283, 202505284 & 3 \\
\hline
\multirow{3}{*}{T2} & No issue & 202505281, 202505282 (accelerometer, gyroscope), 202505283, 202505284 & 4 \\
\cline{2-4}
 & Missing heartRate file & 202505282 & 1 \\
\cline{2-4}
 & Missing orientation file & 202505282 & 1 \\
\hline
\multirow{3}{*}{T4} & No issue & 202505281, 202505283 (accelerometer, gyroscope), 202505284 & 3 \\
\cline{2-4}
 & Missing heartRate file & 202505283 & 1 \\
\cline{2-4}
 & Missing orientation file & 202505283 & 1 \\
\hline
\multirow{1}{*}{T5} & No issue & 202505282, 202505283, 202505284 & 3 \\
\hline
\end{tabularx}

    \caption{Summary of data issues detected in smartwatch recordings from group presentations. If a presentation ID does not appear for a given role, this indicates that the student did not provide consent. In the case of the teacher and observer roles, absence of a presentation ID indicates that the smartwatch was not worn during that presentation}
    \label{tab:available_data_group}
\end{table}

\begin{table}[ht]
    \centering

\begin{tabularx}{\textwidth}{|l|l|X|r|}
\hline
\textbf{Role} & \textbf{Data Issues} & \textbf{ID} & \textbf{Count} \\
\hline
\multirow{4}{*}{E1} & 1-2 min data loss & 202503251 (heartRate, orientation) & 1 \\
\cline{2-4}
 & No issue & 202503112, 202503182, 202503183, 202503311, 202503312, 202503313, 202504011 (gyroscope, orientation), 202504013, 202504081, 202504221, 202504222, 202504223, 202505051, 202505054, 202505061, 202505131 & 16 \\
\cline{2-4}
 & Missing accelerometer file & 202504011 & 1 \\
\cline{2-4}
 & Missing heartRate file & 202504011 & 1 \\
\hline
\multirow{3}{*}{E2} & 1-2 min data loss & 202503182 (accelerometer, heartRate, orientation) & 1 \\
\cline{2-4}
 & No issue & 202503111, 202503171, 202503183, 202503241, 202503251, 202503252, 202503253, 202503311, 202503312, 202504011 (gyroscope, heartRate, orientation), 202504071, 202504072, 202504221, 202504222, 202504223, 202505054, 202505061, 202505062, 202505063, 202505131, 202505134 & 21 \\
\cline{2-4}
 & Missing accelerometer file & 202504011 & 1 \\
\hline
\multirow{6}{*}{O} & 1-2 min data loss & 202503171 (all), 202503181 (gyroscope) & 2 \\
\cline{2-4}
 & No issue & 202503173 (accelerometer, orientation), 202503183 (accelerometer), 202503241, 202503242, 202503243, 202503251, 202503253, 202503311, 202503312, 202503313, 202504011, 202504071, 202504072, 202504081 (accelerometer, gyroscope, orientation), 202504221 (heartRate), 202504222, 202505051 (heartRate), 202505054 (accelerometer, heartRate), 202505131, 202505134 & 20 \\
\cline{2-4}
 & Missing accelerometer file & 202503181, 202504221, 202505051 & 3 \\
\cline{2-4}
 & Missing gyroscope file & 202503173, 202503183, 202504221, 202505051, 202505054 & 5 \\
\cline{2-4}
 & Missing heartRate file & 202503173, 202503181, 202503183, 202504081 & 4 \\
\cline{2-4}
 & Missing orientation file & 202503181, 202503183, 202504221, 202505051, 202505054 & 5 \\
\hline
\multirow{3}{*}{P} & 1-2 min data loss & 202503173 (accelerometer, heartRate, orientation) & 1 \\
\cline{2-4}
 & No issue & 202503111, 202503113, 202503172, 202503241, 202503242, 202503243, 202503251, 202503252, 202503253, 202503311, 202503312, 202503313, 202504011, 202504013, 202504071, 202504072, 202504081, 202504221, 202504222, 202505051, 202505061, 202505063, 202505131, 202505134 & 24 \\
\cline{2-4}
 & Missing gyroscope file & 202503173 & 1 \\
\hline
\multirow{4}{*}{T} & No issue & 202503111, 202503112, 202503113, 202503171, 202503242, 202503243, 202503251, 202503252, 202503253, 202503311, 202503312, 202505054 (accelerometer), 202505061 (accelerometer, gyroscope, orientation), 202505062 (accelerometer, gyroscope, orientation), 202505063, 202505131 & 16 \\
\cline{2-4}
 & Missing gyroscope file & 202505054 & 1 \\
\cline{2-4}
 & Missing heartRate file & 202505054, 202505061, 202505062 & 3 \\
\cline{2-4}
 & Missing orientation file & 202505054 & 1 \\
\hline
\end{tabularx}

    \caption{Summary of data issues detected in smartwatch recordings from individual presentations. If a presentation ID does not appear for a given role, this indicates that the student did not provide consent. In the case of the teacher and observer roles, absence of a presentation ID indicates that the smartwatch was not worn during that presentation}
    \label{tab:available_data_individual}
\end{table}

\subsection*{Eye-tracking Data Quality and Completeness} Eye-tracking data quality and completeness were assessed across the 39 SOPHIAS presentations for which participants consented to being recorded with the Tobii Pro Glasses 3. Before each recording, the Tobii Pro Glasses 3 were calibrated following the standard Tobii procedure. Calibration quality was checked procedurally during acquisition by verifying in the live view that the estimated gaze position was aligned with the participant's fixation target before starting the recording. Because the released Tobii Pro Glasses 3 files do not include a numerical calibration-quality score, post-recording indicators, including valid-sample percentage and missing-eye-data percentage, were used to assess eye-tracking completeness and recording quality.

In total, these recordings contained 1,879,967 gaze samples, of which 1,845,553 were valid, corresponding to an overall valid-sample rate of 98.17\%. At the presentation level, the mean percentage of valid samples across full recordings was 98.24\% (SD = 0.79; range = 95.87–99.35\%), indicating a low average missing-sample rate. Missingness affecting at least one eye was also limited: across the full recordings, the mean percentage of samples with missing data for any eye was 3.53\% (SD = 1.32; range = 1.30–7.58\%). 

During the oral presentation segment, excluding the question-and-answer period, the mean percentage of samples with missing data for any eye was 2.84\% (SD = 1.02; range = 0.99–5.49\%) while the mean percentage of valid samples was 97.16\% (SD = 1.02; range = 94.51–99.01\%).

\subsection*{Audio Synchronization Quality}

Audio synchronization quality was assessed by comparing the ambient audio recorded through Microsoft Teams with the webcam audio captured by edBB. The edBB webcam-audio stream was used as the reference signal because it was recorded within the same infrastructure as the classroom webcam videos and shares the same timestamping framework. It therefore provides the closest internal audio reference for aligning the external ambient microphone with the edBB audiovisual recordings.

For each presentation session, the temporal offset between the Teams audio and the edBB webcam-audio waveforms was estimated using an FFT-based cross-correlation procedure, as described in the Data Processing section; given the relatively short duration of the sessions, no separate time-varying drift correction was applied. After applying the estimated offset correction, the residual synchronization error was obtained by cross-correlating the synchronized Teams audio with the edBB webcam-audio reference and converting the resulting absolute lag to milliseconds. This validation was completed for 48 of the 50 presentation sessions. In the remaining two sessions, the edBB webcam-audio file was not available due to consent restrictions affecting the corresponding webcam-based recording.

The number of valid audio synchronization cases does not necessarily match the number of available webcam video recordings. In some sessions, the webcam video stream is absent due to participant consent restrictions, while the associated webcam-audio stream is still available and can be used for synchronization assessment. Thus, the 48 valid cases reported here refer specifically to sessions for which both Teams audio and edBB webcam audio were available.

Across the 48 valid sessions, the residual absolute synchronization error after correction had a mean of 7.08 ms (SD = 10.06), a median of 1.13 ms, and a maximum of 35.00 ms. All residual errors were below the duration of one frame in the 20 Hz edBB webcam video recordings (50 ms), indicating sub-frame alignment of the Teams audio with the edBB webcam-audio reference in all valid cases.

\subsection*{Smartwatch Heart-Rate Data Availability and Completeness}

Heart-rate data availability and completeness were assessed for the Fitbit Sense smartwatch recordings available in SOPHIAS. Across all available smartwatch heart-rate recordings, SOPHIAS contains 147,640 heart-rate samples. Of these, 80,865 samples correspond to the oral presentation segments, while the remaining 66,775 samples correspond to the question-and-answer periods.

To assess heart-rate completeness, we computed the percentage of valid heart-rate samples for each available role-specific recording. For peer evaluator E1, the mean valid-sample percentage was 91.18\% (SD = 11.08). For peer evaluator E2, the mean was 93.49\% (SD = 8.24). For the external observer, the mean was 94.25\% (SD = 13.85). For presenters, the mean was 90.95\% (SD = 10.65). For teachers, the mean was 85.66\% (SD = 12.31).

\subsection*{Face Extraction and Head Pose Output Coverage}

We validated the face-extraction pipeline applied to the available video streams to assess the completeness and consistency of the visual features obtained from the recordings. Face detection was performed using RetinaFace, which provided the face bounding boxes and facial landmarks used as input for the subsequent head-pose estimation module.

For the presenter-view webcam stream, we analyzed the 38 available videos corresponding to individual presentations. Across these videos, the pipeline processed a total of 753,667 frames. RetinaFace successfully detected a face in 736,126 frames, corresponding to a frame-weighted global detection rate of 97.67\%. The remaining 17,541 frames, representing 2.33\% of the total, did not contain a valid face detection. At the video level, the mean face-detection rate was 97.92\% (SD = 3.86 percentage points), indicating consistently high coverage across the available presenter-view recordings.

We also validated face extraction on the eye-tracker scene videos. We analyzed the 37 available recordings. Across the included eye-tracker scene videos, the pipeline processed 883,007 frames. A valid face detection was obtained in 872,959 frames, corresponding to a frame-weighted global detection rate of 98.86\%. The remaining 10,048 frames, representing 1.14\% of the total, did not contain a valid face detection. At the recording level, the mean detection rate was 98.93\% (SD = 1.33 percentage points).

In addition, all frames with a valid face detection also had an associated head-pose estimation. Therefore, the coverage of the head-pose estimation module over the face-detected frames was 100\%, ensuring that every valid face crop included the corresponding yaw, pitch, and roll values.

\begin{figure}[th]
    \centering
    \includegraphics[width=0.95\linewidth]{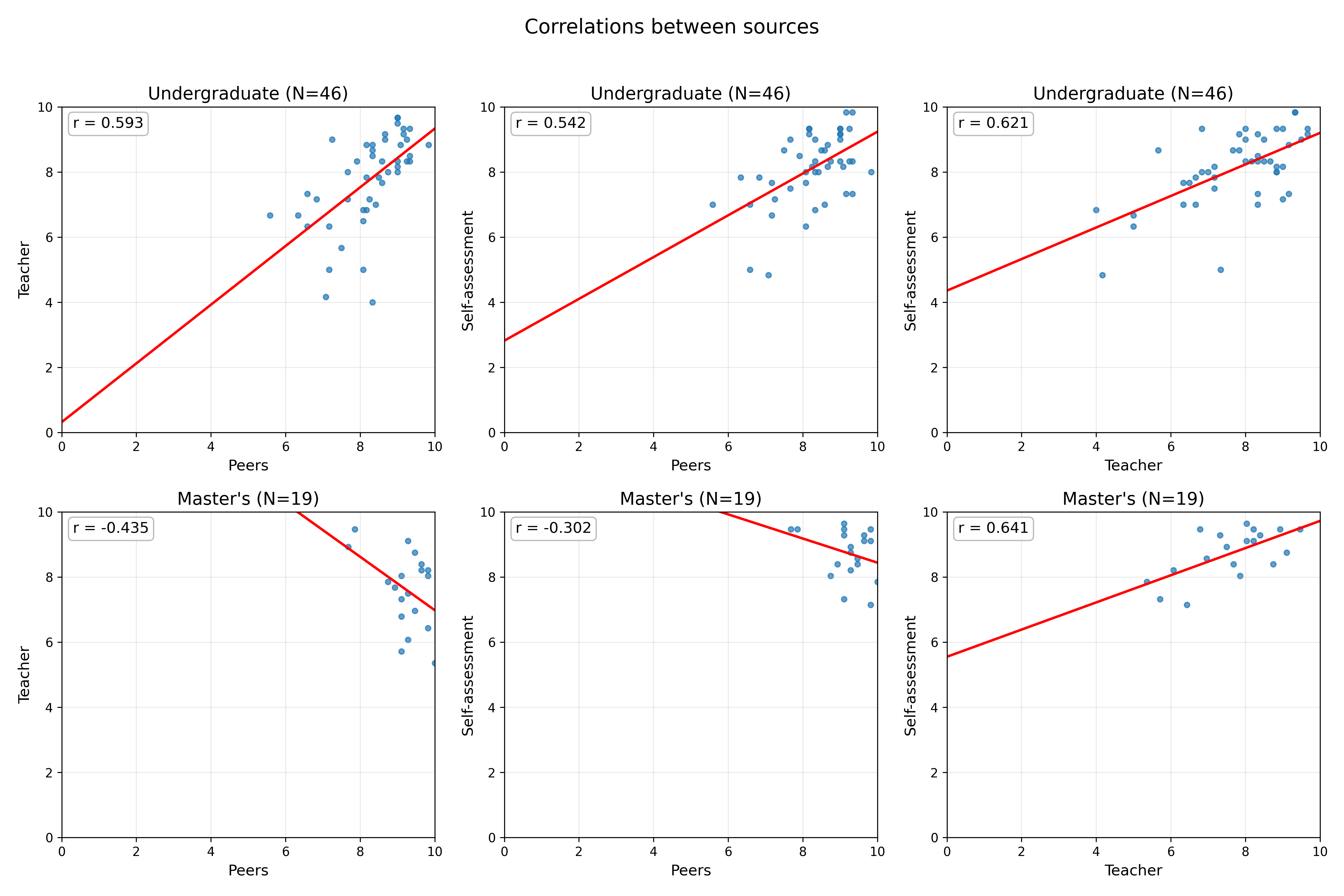}
    \caption{Pairwise correlations between grades across evaluators (teacher, peers, and self-assessments). Each subplot displays the Pearson correlation coefficient ($r$) and the regression line (in red), with all grades normalized to a 0--10 scale.}
    \label{fig:correlations}
\end{figure}

\subsection*{Teacher, Peer and Self-Assessment Validation}
The SOPHIAS dataset incorporates an evaluation methodology involving teacher, peer, and self-assessments through the AICoFe system. To validate the reliability of these assessments, we analyzed the level of agreement for both undergraduate ($N_{\text{students}} = 46$) and master’s ($N_{\text{students}} = 19$) presentations using Gwet’s AC2 coefficient \cite{vach2023gwet}  with quadratic weights, with 95\% confidence intervals calculated using the Student's $t$ distribution.. We used Gwet's AC2 as the primary inter-rater reliability statistic because the 
rubric scores are ordinal Likert-type ratings and several items showed highly 
skewed marginal distributions. Under these conditions, Cohen's $\kappa$ can be 
unstable and may produce artificially low or negative values despite substantial 
observed agreement, a limitation commonly referred to as the kappa paradox. 
Gwet's AC2 is less sensitive to prevalence and marginal imbalance and is 
therefore more appropriate for the present assessment data. The intraclass 
correlation coefficient (ICC) was not used because the assessment design was not 
fully crossed: peer evaluators varied across presentations, rather than the same 
raters evaluating all presenters. Moreover, teacher, peer, and self-assessments 
represent different assessment sources, not equivalent raters drawn from a 
common rater population. Agreement was therefore computed using Gwet's AC2 for 
the teacher--peer, peer--self, and teacher--self comparisons, as well as 
globally across all available assessment sources. No incomplete rubric scores were present in the assessment data; therefore no imputation was required.

Additionally, we examined the correlations among the different evaluative sources using Pearson’s correlation coefficient (r) and Spearman's correlation ($\rho$) and investigated potential scoring biases (whether any group of evaluators tended to assign systematically higher or lower scores) through paired-sample tests (Student’s t-test or Wilcoxon signed-rank test, depending on normality assumptions) \cite{becerra2025leveraging}.

These analyses were conducted exclusively for technical validation purposes, with the aim of assessing the internal consistency, coherence, and plausibility of the captured evaluation data rather than testing instructional or behavioral hypotheses.
All statistical analyses were conducted using Python. Data preparation, aggregation, and correlation analyses were performed in Python using \texttt{pandas} for data handling and \texttt{scipy.stats} for Pearson and Spearman correlation coefficients. Inter-rater agreement analyses based on Gwet’s AC2 were conducted using the \texttt{irrCAC} package.

\begin{figure}[ht]
    \centering
    \includegraphics[width=\linewidth]{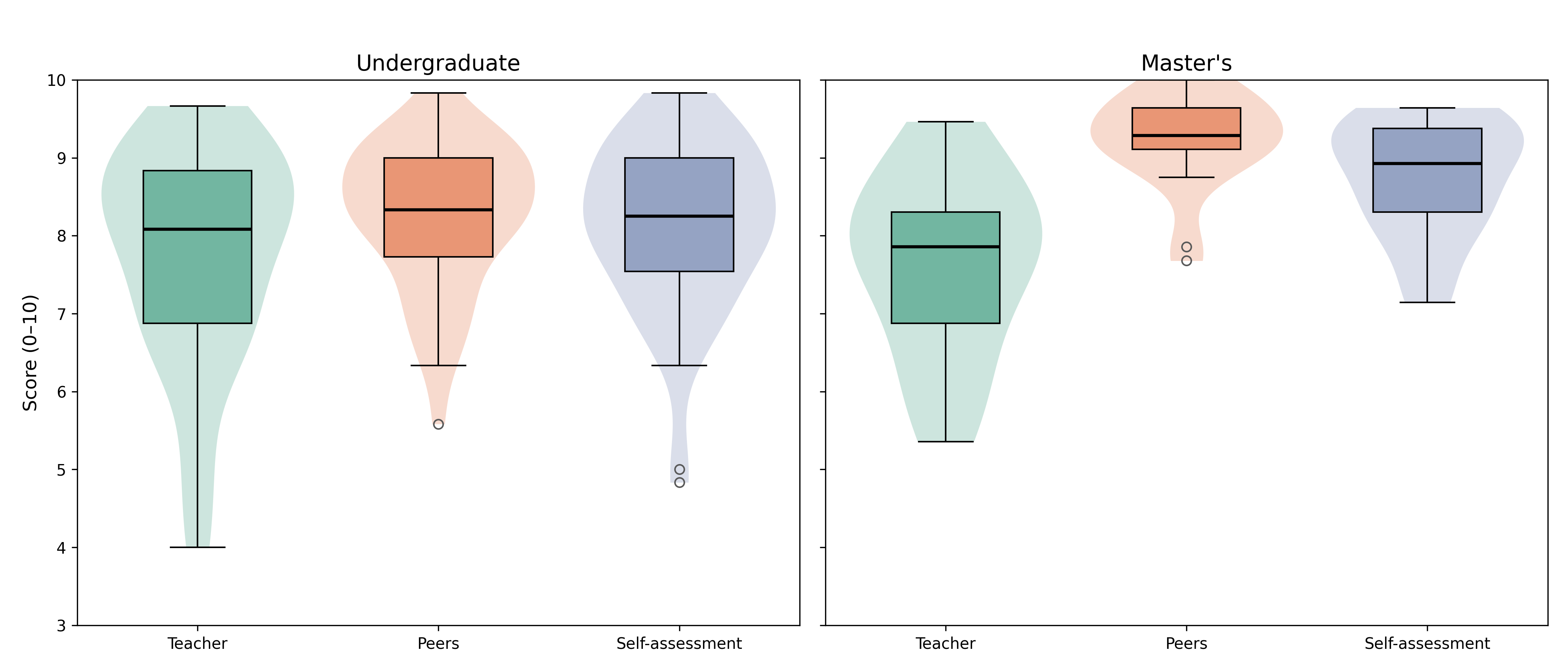}
    \caption{Distribution of scores across evaluation sources. All scores were normalized to a 0--10 scale.}
    \label{fig:bias}
\end{figure}

For the undergraduate presentations, mean pairwise comparisons showed moderate to substantial agreement for teacher vs peers (AC2 = 0.60, 95\% CI [0.53, 0.68]), peers vs self-assessment (AC2 = 0.63, 95\% CI [0.56, 0.71]), and teacher vs self-assessment (AC2 = 0.58, 95\% CI [0.51, 0.65]). According to established interpretation thresholds, AC2 values between 0.40 and 0.60 indicate moderate agreement, while values above 0.60 are considered substantial. When considering all evaluators together, global agreement reached AC2 = 0.66 (95\% CI [0.63, 0.70]), confirming consistent agreement patterns across sources.

For the master's presentations, agreement levels showed differences compared with undergraduates. The mean agreement between teacher and peer evaluations was moderate (AC2 = 0.49, 95\% CI [0.35, 0.64]), while peer vs self-assessment reached a substantially higher level (AC2 = 0.70, 95\% CI [0.59, 0.81]). In contrast, teacher vs self-assessment showed lower concordance (AC2 = 0.55, 95\% CI [0.40, 0.69]), indicating that, in terms of agreement coefficients, master's students' self-assessments were more closely aligned with peer evaluations than with teacher evaluations. When considering all evaluators together, global agreement reached AC2 = 0.65 (95\% CI [0.55, 0.74]).

These values, ranging from moderate to substantial agreement, are consistent with previous findings in peer assessment research, which typically report stronger alignment among student-driven evaluations and lower concordance between teacher and self-assessments \cite{tang2023comparison,ross2006reliability,kilic2016examination}.

The correlation analyses offer further evidence of the validity and coherence of the evaluation process (Figure~\ref{fig:correlations}). For undergraduate presentations, the strongest association was found between the teacher’s and self-assessment rubric means ($r = 0.62$, 95\% CI [0.40, 0.77], \(\rho = 0.56\)), followed by teacher–peers ($r = 0.59$, 95\% CI [0.37, 0.75], \(\rho = 0.69\)) and peers–self ($r = 0.54$, 95\% CI [0.30, 0.72], \(\rho = 0.49\). These values are comparable to the average correlation reported in large-scale meta-analyses of peer and teacher assessments, which range from moderate to substantial across diverse educational contexts \cite{falchikov2000student,li2016peer}.

For master's presentations, the pattern was notably different: the highest correlation was observed between teacher and self-assessment ($r = 0.64$, 95\% CI [0.26, 0.85], \(\rho = 0.57\)), while the relationship between peers and self-assessment was negative ($r = -0.30$, 95\% CI [-0.67, 0.18], \(\rho = -0.22\)), and teacher–peer correlation was also negative ($r = -0.44$, 95\% CI [-0.74, 0.02], \(\rho = -0.26\)). 
This divergence indicates that master's students’ self-assessments align more closely with teacher evaluations than with peer scores. In fact, peer evaluations tended to be consistently generous, regardless of actual performance, which suggests a leniency or benevolence bias aimed at supporting classmates rather than strictly applying the rubric. Several studies point to systematic biases in peer evaluation, such as leniency, friendship, or reciprocity effects, which can undermine reliability \cite{magin2001reciprocity,stonewall2018review}. While \cite{magin2001reciprocity} demonstrated that reciprocity effects account for only a small portion of score variance, later studies highlight the persistence of biases, particularly in collaborative and team-based learning environments \cite{stonewall2018review}.

Regarding scoring biases (Figure \ref{fig:bias}), for undergraduate students, self-assessments were significantly higher than teacher scores ($t(45) = 2.15, p = 0.037;$ mean difference $= +0.36$ points), while peers also scored higher than teachers (Wilcoxon $W = 267.50, p = 0.008;$ mean difference $= +0.49$ points). However, the difference between peers and self-assessments was not significant ($t(45) = -0.93, p = 0.356;$ mean difference $= -0.14$ points). For master's students, teacher scores were significantly lower than those assigned by peers ($t(18) = -4.56, p < 0.001;$ mean difference $= -1.60$ points) and self-assessments ($t(18) = -5.44, p<0.0001;$ mean difference $= -1.11$ points). In contrast, the difference between peers and self-assessments was not statistically significant ($t(18) = 1.93, p = 0.07$), although peers tended to score slightly higher ($+0.49$ points).

\begin{figure}[t]
    \centering
    \includegraphics[width=\linewidth]{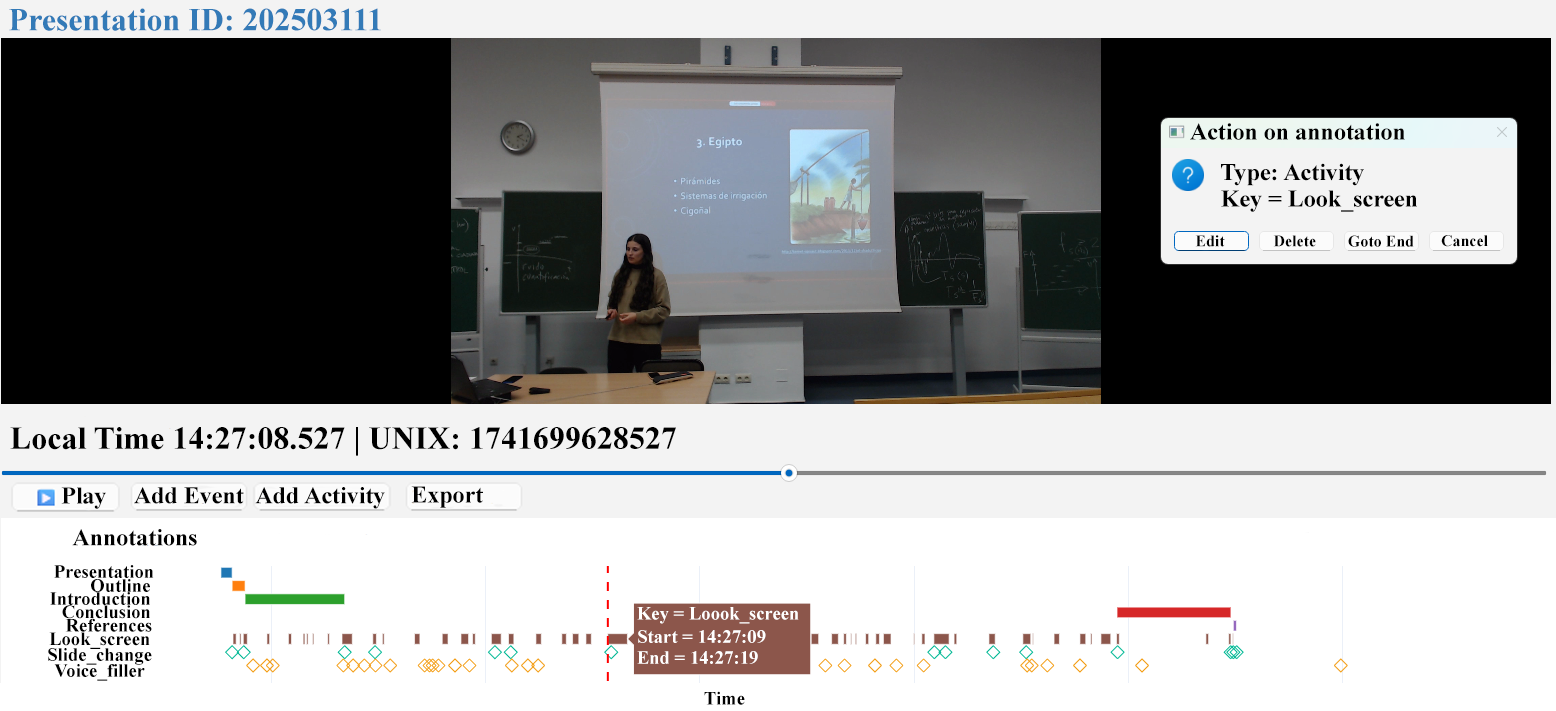}
\caption{Screenshot of the HuMLaS graphical user interface used for manual 
review and refinement of behavioral annotations. The interface allows annotators 
to inspect synchronized video recordings from Logitech C920 PRO HD webcam 
videos recorded at 20 Hz and to adjust event boundaries using frame-level 
timestamps. The participant shown in the screenshot provided consent for 
publication of this image.}
    \label{fig:humlas}
\end{figure}

These findings align with previous research showing that both undergraduate and master's students tend to rate themselves and their peers slightly higher than teachers, reflecting a modest positive scoring bias consistent with prior evidence that raters can systematically overestimate performance \cite{oren2018self,falchikov2000student,patri2002influence}.

\subsection*{Human Re-labeling with HuMLaS}

To ensure high-quality, temporally precise, and semantically consistent annotations, we incorporated an additional human-in-the-loop validation step using HuMLaS (Human Labeling System). HuMLaS (Figure \ref{fig:humlas}) is a system that allows annotators to manually review video recordings, inspect the associated behavioral events, and refine labels directly on an interactive timeline.

Each video in SOPHIAS was recorded with per-frame timestamps. HuMLaS leverages these timestamps so that annotators can accurately adjust the start frame at which a behavior begins, the end frame at which it concludes, and therefore the exact temporal interval associated with each label.

Within the interface, annotators can scrub through the video frame by frame, align the event markers with visible behavioral cues (e.g., the first moment of looking at the computer screen, onset of reading notes, slide changes), and update the annotation boundaries. HuMLaS automatically maps these corrected boundaries back to precise timestamps, ensuring that each label corresponds to an exact segment in the video stream. This process makes the final annotations significantly more reliable than the initial real-time judgments.

Initially, SOPHIAS included a base set of verbal and non-verbal behavioral labels: Confident, Eye\_contact, Stressed, Static, Natural\_gestures, Interaction, Reading (from projection, computer screen, or notes), Nervous\_movement, Filler\_word, Question, Answer, Clear\_voice, and Unclear\_voice. These labels were manually annotated in real time while the presentations were being delivered, providing the first layer of behavioral information, though they were subject to noise and inconsistencies inherent to real-time labeling.

During the technical validation phase, HuMLaS was used to expand, refine, and correct this initial annotation layer. Annotation refinement was performed by a single trained annotator using HuMLaS and all available videos were reviewed by this annotator. To improve consistency, the annotator followed a predefined annotation guideline specifying the meaning of each label, examples of observable behaviors, and rules for assigning start and end times. The human annotator identified behaviors that were missing, underspecified, or semantically ambiguous, and incorporated new distinctions to better capture the structure and delivery of each presentation. 

During the human re-labeling process, the original annotation set was revised following a structured approach. 
First, high-level labels were introduced to capture the structural organization of the presentations. 
Second, several existing labels were refined or disambiguated to improve semantic clarity and consistency.

All original labels remain available in the dataset. The list below reports the labels that were newly introduced or explicitly redefined during this refinement process:

\begin{itemize}
    \item \textbf{Structural segments of the presentation (added):} 
    Presentation (opening), Introduction, Outline, Conclusion, References, End\_presentation.

    \item \textbf{Reading-related behaviors (refined):} 
    Look\_screen (looking at the computer screen), Read\_notes (reading from paper notes, mobile phone, or other written material).

    \item \textbf{Movement-related behaviors (refined):} 
    Nervous\_movement (refined from the original category to improve temporal precision and consistency).

    \item \textbf{Expressiveness levels (added):} 
    Nervous, Enthusiastic, Flat.

    \item \textbf{Slide transition events (added):} 
    Slide\_change, Slide\_animation.

    \item \textbf{Verbal disfluency labeling (refined):} 
    Voice\_filler is the label used in the HuMLaS re-labeling to refer to the pre-existing \textit{Filler\_word} category, which was annotated in real time during the presentations. Both labels denote the same phenomenon (verbal fillers).

    \item \textbf{Q\&A-related events (reviewed and corrected):} 
    Question, Answer.

    \item \textbf{Multimedia usage during the presentation (added):} 
    Video\_use.
\end{itemize}

\section*{Usage Notes}
\textbf{Synchronization across modalities:}
For synchronization, users should rely on the UNIX timestamp fields rather than local-time strings. Event-based data, such as annotations, keyboard, mouse, and clicker logs, can be aligned directly to continuous streams by matching their UNIX timestamps to the nearest sample or video frame timestamp. Continuous signals with different sampling rates, such as eye-tracking data, video, audio, and smartwatch signals, can be synchronized by converting UNIX timestamps to a common millisecond time axis and then resampling, interpolating, or windowing the signals according to the temporal resolution required by the analysis. Thus, all modalities can be temporally aligned using UNIX time in milliseconds as the shared reference clock, while local timestamps are provided primarily for human readability and inspection.

\textbf{Units of analysis:} SOPHIAS can be analyzed at several levels depending on the research question. The student is the participant-level unit and is identified by an internal AICoFe ID. The presentation is the session-level unit and is identified by the presentation folder name. The role is a session-specific unit indicating each participant's function within a given presentation, such as \texttt{P}, \texttt{T}, \texttt{O}, \texttt{E1}, and \texttt{E2} for individual presentations, or indexed roles such as \texttt{P1}--\texttt{P5}, \texttt{T1}--\texttt{T5}, and \texttt{E1}--\texttt{E5} for group presentations. For master's presentations, the group is an additional unit of analysis, since each presentation was delivered by a group of four or five students. Evaluation data can also be analyzed by evaluation source, distinguishing teacher, peer, and self-assessments. Users should explicitly report the unit of analysis used in each study, especially when combining modalities or comparing individual and group presentations.

\textbf{Suggestions:} The \textbf{Pandas} library (\url{https://pandas.pydata.org/}) proved very useful for manipulating and analyzing large datasets, while \textbf{NumPy}\cite{harris2020array} (\url{https://numpy.org}) was used for processing various types of signals and videos. For training deep learning models or neural networks, \textbf{TensorFlow}\cite{abadi2016tensorflow} (\url{https://www.tensorflow.org/}) and \textbf{PyTorch}\cite{paszke2019pytorch} (\url{https://pytorch.org/}) are two of the most powerful and flexible platforms for this purpose, and were essential for running the models used in video processing.

For video processing, the \textbf{Dlib}\cite{king2009dlib} library (\url{www.dlib.net}), which includes pre-trained models for facial detection, facial landmark estimation, and more, was used. This library integrates well with \textbf{OpenCV} (\url{https://opencv.org/}), which facilitates video reading, filtering, distortion correction, and also provides additional pre-trained models. This combination was used to process the videos, alongside \textbf{InsightFace}\cite{deng2018arcface} (\url{https://github.com/deepinsight/insightface}), a framework that specializes in models for face detection. The use of \textbf{RetinaFace}\cite{deng2020retinaface}  (\url{https://github.com/serengil/retinaface}) is also recommended for facial detection with landmark localization. 

 \textbf{WHENet}\cite{zhou2020whenet} (\url{www.github.com/Ascend-Research/HeadPoseEstimation-WHENet}) is a fast and efficient method for head pose estimation, which was used to obtain the Euler angle labels in the SOPHIAS dataset. Another effective alternative for head pose estimation is \textbf{RealHePoNet} \cite{berral2021realheponet}  (\url{https://github.com/rafabs97/headpose_final}).

For processing eye-tracker data, the official software \textbf{Tobii Pro Lab} (\url{https://www.tobii.com/products/software/behavior-research-software/tobii-pro-lab}) is recommended. Another interesting solution is \textbf{PsychoPy}\cite{peirce2007psychopy} (\url{https://www.psychopy.org/}), a free tool offering support for the creation and execution of experiments involving eye tracking.

\textbf{Limitations and recommended use:} SOPHIAS was collected in authentic classroom settings and therefore prioritizes ecological validity over experimental control, in line with recent calls for MMLA research to move beyond laboratory settings and toward authentic learning environments \cite{schneider2026implementing}. The following considerations are intended to guide appropriate reuse of the dataset.

First, the dataset includes 50 oral presentations delivered by 65 students from a single university and from engineering- and data-science-related courses. Consequently, SOPHIAS should not be interpreted as representative of all higher-education contexts, disciplines, cultural settings, or student populations. Researchers interested in generalizable claims should treat SOPHIAS as an ecologically valid case study dataset and, where possible, complement it with data from additional institutions, disciplines, or cohorts.

Second, the undergraduate and master’s subsets differ in structure and sample size. Undergraduate presentations were individual, whereas master’s presentations were group-based. These differences provide opportunities to study both individual and group presentation settings, but they should be considered when pooling data across degree levels or comparing undergraduate and master’s presentations. We recommend that users explicitly report which subset they analyze and avoid merging both subsets without accounting for differences in presentation format, assessment structure, and participant roles.

Third, the sample presents demographic imbalance, particularly in gender distribution. This imbalance may affect analyses focused on individual differences, fairness, subgroup comparisons, or model generalization, especially because prior MMLA research has shown that gender and group composition can shape multimodal participation dynamics in collaborative learning contexts \cite{lee2026toward}. Researchers should therefore report demographic inclusion criteria, avoid drawing strong conclusions for underrepresented groups, and consider demographic variables when evaluating model performance or assessment patterns. For predictive modeling, mitigation strategies may include participant-independent splits that preserve subgroup representation where feasible, subgroup-specific performance evaluation, sample or loss weighting, balanced sampling or controlled over-sampling of underrepresented groups within the training set, and modality-appropriate data augmentation applied exclusively to the training folds to avoid data leakage.

Fourth, data availability varies across modalities because participants could choose which sensors and recordings they consented to provide. As a result, some presentations lack specific video, eye-tracking, interaction log, or smartwatch streams. This situation corresponds to a missing-modality scenario, a common challenge in multimodal learning settings where one or more modalities may be unavailable because of sensor limitations, privacy constraints, cost, or data loss \cite{wu2024deep}. The smartwatch modality has the highest level of missingness, reflecting both consent-related limitations and occasional technical issues. Users should inspect the missingness tables and metadata before analysis, select subsets according to the modalities required by their research question, and clearly report the inclusion and exclusion criteria used to construct their analytical sample.

Fifth, missing sensor data may affect analyses that depend on complete multimodal coverage, especially team-coordination analyses in group presentations. For example, missing physiological or inertial data for specific roles may prevent the construction of complete interaction, synchrony, or coordination networks. In these cases, researchers may restrict analyses to roles and modalities with sufficient coverage, model missingness explicitly, or use methods designed for incomplete multimodal data. Missing nodes or streams should be reported transparently when interpreting coordination results.

Sixth, the contextual annotation layer was refined using HuMLaS by a single trained annotator following predefined annotation guidelines. Although this procedure improved temporal precision and semantic consistency compared with the original real-time annotations, inter-annotator agreement could not be computed. Therefore, the behavioral labels should be treated as carefully curated annotations rather than as labels with independently validated inter-rater reliability. At the same time, the released annotations lower the entry cost for secondary users by providing pre-segmented and timestamped candidate events. Depending on their research aims, users may reuse these labels as provided, validate a subset independently, adapt the annotation scheme, or perform additional re-annotation using the raw videos and synchronized timestamps. For high-stakes supervised machine-learning tasks or consequential predictive applications in which these behavioral labels are used as target variables, researchers should independently re-annotate a representative subset using at least two trained annotators and compute and report an appropriate inter-annotator agreement measure before treating the labels as validated ground truth.

Seventh, the teacher, peer, and self-assessment scores should be interpreted as complementary evaluation sources rather than interchangeable measures of presentation quality. The technical validation revealed moderate to substantial agreement in several comparisons, but also systematic differences between sources. In particular, peer and self-assessments may reflect leniency, generosity, self-perception, or role-related effects. For this reason, the released scores are provided in their original form. Users interested in predictive modeling, assessment fairness, or evaluator behavior may apply normalization, rater-bias modeling, or other correction procedures appropriate to their specific research questions, while clearly distinguishing between raw and transformed scores.

Eighth, physiological data collected with Fitbit Sense smartwatches should be interpreted with the usual caution associated with wrist-worn photoplethysmography and inertial sensing in real-world settings. Heart rate and movement signals may be affected by device placement, motion artifacts, short data losses, and the low sampling frequency available in the released streams. These signals are suitable for exploratory multimodal analyses and coarse temporal patterns, but users should apply appropriate signal-quality checks before using them for fine-grained physiological inference.

Ninth, SOPHIAS contains sensitive human-subject data, including faces, voices, gaze information, physiological signals, demographic variables, assessment scores, and textual comments. For this reason, access to the full dataset is controlled through a Data Usage Agreement. This access model protects participant privacy while still enabling legitimate research reuse. Users should follow the DUA terms, avoid re-identification attempts, avoid redistribution of raw or derived data, and ensure that any released models, features, or examples do not compromise participant confidentiality.

Tenth, processed visual features such as face detections, facial landmarks, and head-pose estimates should be interpreted as model-derived features rather than ground-truth measurements. Users requiring different detection thresholds, model versions, or feature definitions may reprocess the raw videos using their own pipelines.

Eleventh, because SOPHIAS integrates streams with different sampling rates and acquisition mechanisms, users should verify the temporal alignment assumptions required by their specific analyses, particularly when combining frame-level video features, audio, eye-tracking, smartwatch signals, and interaction events \cite{chango2022review}.

Finally, SOPHIAS combines a moderate number of participants with high temporal and multimodal resolution, reflecting a common trade-off in ecological MMLA: richer in-situ sensing increases authenticity and temporal detail, while also introducing challenges related to scalability, noise, missingness, synchronization, and privacy \cite{schneider2026implementing}. Accordingly, although the participant-level sample size should be considered when making population-level or generalization claims, SOPHIAS contains a substantial volume of reusable multimodal observations across approximately 12 hours of synchronized classroom recordings. The dataset is therefore especially suitable for analyses that exploit temporal granularity, multimodal alignment, behavioral sequences, and presentation-level patterns. For machine-learning applications, all samples or temporal segments associated with the same student or presentation should remain within the same training, validation, or test partition to prevent data leakage. Group presentations require additional care because multiple participants occur within the same session; consequently, session and participant membership should both be considered when constructing evaluation partitions. Researchers should also report the level of analysis, the modalities included, and the completeness and demographic composition of each partition. In downstream predictive studies involving limited or imbalanced analytical subsets, researchers may consider task- and modality-specific strategies such as training-set-only data augmentation, feature selection, regularization, or sample weighting.

\textbf{Data Access}:

The SOPHIAS dataset is available for research use only, including both academic and legitimate commercial research and development. In this context, commercial research and development refers to research activities conducted by commercial organizations, such as internal feasibility studies, algorithm development, or algorithm benchmarking, under the same controlled-access conditions and Data Usage Agreement applicable to academic researchers. Due to the inclusion of sensitive human-subject data, including biometric signals and indirect identifiers (e.g., gender, age), access to the dataset is restricted and reuse is governed by the Data Usage Agreement (DUA) associated with the controlled-access repository, rather than by an unrestricted open data license. The DUA prohibits redistribution of the data in any form (e.g., original files, encrypted files, or extracted features). Facial blurring was not applied to the raw MP4 recordings because facial expressions, gaze direction, head pose, and body posture are core research signals in SOPHIAS. Applying face anonymization would substantially reduce the utility of the videos for multimodal learning analytics and automated feedback research. Data access will be granted to any applicant who agrees to the terms and conditions outlined in the DUA.
Researchers can request access through two available methods. First, the DUA is available on the project’s GitHub
(\url{https://github.com/dataGHIA/SOPHIAS}); a signed and scanned copy must be emailed to ghia.uam@gmail.com, following
the instructions provided in the repository. Alternatively, access requests can be submitted via the Science Data Bank
(\url{https://doi.org/10.57760/sciencedb.33655}), where detailed application procedures are provided.

The current SOPHIAS deposit should be regarded as a static release comprising the 50 presentation sessions described in this manuscript. No periodic schedule has been established for appending additional cohorts to this deposit. Although further cohorts may be considered for publication in the future, any such data would be subject to the corresponding ethical approval, participant consent, privacy review, curation, and technical validation.

\section*{Code Availability}
The code for filtering physiological and biometric signals is available on GitHub (\url{https://github.com/dataGHIA/SOPHIAS}). Python was used for signal filtering and all necessary packages are listed in the requirements.txt file. The GitHub repository also includes an example Jupyter notebook demonstrating how to synchronize smartwatch heart-rate (HR) data with the timestamped contextual annotations using the shared time reference, as well as the machine-readable JSON Schema files and the
dataset catalog.

\section*{Acknowledgements}
Support by projects: Cátedra ENIA UAM-VERIDAS en IA Responsable (NextGenerationEU PRTR TSI-100927-2023-2), M2RAI (PID2024-160053OB-I00, MICIU/FEDER) and SNOLA (RED2022-134284-T). Alvaro Becerra is funded by a predoctoral contract (FPI) from the Comunidad de Madrid (PIPF-2024/COM-34288).

The authors thank Pablo Villegas for wearing the eye-tracker glasses and helping deploy the smartwatches using Watch-DMLT for the SOPHIAS dataset.

\section*{Author Contributions Statement}

\textbf{A. B.:} Conceptualization, Data Collection, Data Curation, Data Analysis, Investigation, Software, Technical Validation, Writing—Original Draft, Writing—Review \& Editing.
\textbf{R. C.:} Data Collection, Funding Acquisition, Investigation, Supervision, Writing—Review \& Editing.
\textbf{R. D.:} Funding Acquisition, Investigation, Software, Supervision, Writing—Review \& Editing.

\section*{Competing interests}
The authors declare no competing interests.

\FloatBarrier

\bibliography{sample}
\end{document}